\documentclass[prl, aps, 10pt, twocolumn, superscriptaddress, showpacs]{revtex4-1}

\usepackage{bbold}
\usepackage{color}
\usepackage{graphicx}
\usepackage{hyperref}
\usepackage{mathtools}
\usepackage{amssymb, amstext, amsmath}

\DeclareMathOperator{\Tr}{Tr}
\DeclarePairedDelimiter\abs{\lvert}{\rvert}

\newcommand{\iu}{\mathrm{i}}
\newcommand{\po}{\widetilde}

\begin{document}

%% =============================================================================
%%
%% TITLE PAGE
%%
\title{Environment mediated multipartite and multidimensional entanglement}

\author{Chee Kong Lee}
\affiliation{Centre for Quantum Technologies, National University of Singapore,
             117543, Singapore}
\affiliation{Department of Chemistry, Massachusetts Institute of Technology,
             Cambridge, Massachusetts 02139, USA}

\author{Mojdeh S. Najafabadi}
\affiliation{Dodd-Walls Centre for Photonic and Quantum Technologies,
             Department of Physics, University of Otago, Dunedin, New Zealand}

\author{Daniel Schumayer}
\affiliation{Dodd-Walls Centre for Photonic and Quantum Technologies,
             Department of Physics, University of Otago, Dunedin, New Zealand}

\author{Leong Chuan Kwek}
% \email{cqtklc@nus.edu.sg}
\affiliation{Centre for Quantum Technologies, National University of Singapore,
             117543, Singapore}
\affiliation{Institute of Advanced Studies, Nanyang Technological University,
             60 Nanyang View Singapore 639673, Singapore}
\affiliation{National Institute of Education, 1 Nanyang Walk Singapore 637616,
             Singapore}
\affiliation{MajuLab, CNRS-UNS-NUS-NTU International Joint Research Unit,
             UMI 3654, Singapore}

\author{David A.~W. Hutchinson}
\email{david.hutchinson@otago.ac.nz}
\affiliation{Centre for Quantum Technologies, National University of Singapore,
             117543, Singapore}
\affiliation{Dodd-Walls Centre for Photonic and Quantum Technologies,
             Department of Physics, University of Otago, Dunedin, New Zealand}
\affiliation{Department of Physics, Clarendon Laboratory, University of Oxford,
              Parks Road, Oxford OXI 3PU, United Kingdom}

\begin{abstract}
Quantum entanglement is usually considered a fragile quantity and decoherence
through coupling to an external environment, such as a thermal reservoir, can
quickly destroy the entanglement resource. This doesn't have to be the case and
the environment can be engineered to assist in the formation of entanglement.
We investigate a system of qubits and higher dimensional spins interacting only
through their mutual coupling to a reservoir. We explore the entanglement of
multipartite and multidimensional system as mediated by the bath and show that
at low temperatures and intermediate coupling strengths multipartite
entanglement may form between qubits and between higher spins, i.e., qudits. We
characterise the multipartite entanglement using an entanglement witness based
upon the structure factor and demonstrate its validity versus the directly
calculated entanglement of formation, suggesting possible experiments for its
measure.
\end{abstract}

%\pacs{03.65.Yz, 71.38.-k}
\date{\today} \maketitle

%% =============================================================================
%% INTRODUCTION
%% =============================================================================
Entanglement between systems with few degrees of freedom is often a fragile
property as their interaction with a much larger environment drives decoherence.
This decoherence often comes from  the small system interacting in an
incoherent, memory-less manner with the bath. Most theoretical studies of
few-body quantum systems make this assumption explicitly or implicitly, i.e.,
the dynamics is Markovian. Such approximation seems quite appropriate and
experimentally valid to high precision in condensed matter
systems~\cite{Gardiner2010}. However, it is also a possibility that the
environment, if suitably prepared, can mediate the creation of quantum
correlation between small systems~\cite{Braun2002}.

Quantum entanglement for two or even three qubits has been studied extensively
\cite{Amico2008, Horodecki2009} and it is well understood, but less well
characterised is {\emph{multipartite entanglement}} -- entanglement between
several systems. For example, for more than four qubits, the creation as well as
the characterisation of quantum entanglement becomes exceedingly difficult.
Theoretical investigations also face the curse of dimensionality, since the
dimension of the corresponding Hilbert-space grows exponentially with the number
of subsystems. Despite the hurdles, experimental demonstration of multipartite
entanglement has been achieved in ion traps with up to 14
ions~\cite{Haeffner2005, Leibfried2005, Monz2011}, or for ten superconducting
qubits~\cite{Vidal2003, Song2017} and ten photonic qubits on a linear optical
platform~\cite{Wang2016}.

The main obstacle to achieving significant multipartite entanglement is
decoherence and dissipation due to coupling to the environment. The lifetime of
entangled states decreases rapidly with the size of the system. Standard
approaches to overcome this difficulty involve high fidelity quantum gates and
large qubit overheads, which can be more and more challenging as the number of
qubits increases. On the other hand, steady state two-qubit entanglement can be
generated by engineered dissipation in trapped ions~\cite{Lin2013} and
superconducting qubits~\cite{Shankar2013}, and noise assisted quantum transport
has also been reported~\cite{Viciani2015, Biggerstaff2016}. These experimental
results challenge the general assumption that dissipation is always detrimental
to quantum information processing, rather they demonstrate that it might be a
resource that can be engineered and harnessed.

It has been conjectured that non-Markovian effects in the quantum time
evolution, assuming an Ohmic bosonic bath at zero temperature, can leave their
fingerprints on the time dependence of coherence~\cite{Divincenzo2005}. This
conjecture relies on the applicability and accuracy of the Born approximation
underpinning the Bloch-Redfield equation. It should be noted that this
perturbation approach contains accumulating divergences \cite{Slutskin2011},
rendering the Born approximation inappropriate at long times. As a consequence
care should be taken when interpreting these theoretical investigations, but the
different Markovian and non-Markovian processes which affect relaxation and
dephasing do offer opportunities to engineer the effects of the environment upon
the system of interest.

Another research direction proposes entangling systems with more numerous
degrees of freedom ~\cite{Mair2001}. In general we may replace a standard
two-level system with a $d$-dimensional system -- a {\emph{qudit}}. Such
generalisation is not a simple extension of already existing concepts: the
entanglement of two-qubit systems seems somewhat the
exception~\cite{Vollbrecht2000} rather than the rule. In addition, there is
qualitative difference between bipartite and multipartite
entanglement~\cite{Bengtsson2016} and the latter may provide higher flexibility
to achieve more efficient quantum information processing~\cite{Groeblacher2006}.
The main focus of recent experiments has been to generate and characterise
multilevel and multipartite entanglement~\cite{Mair2001, Martin2017, Krenn2013,
Molina-Terriza2007}. Our work presented here aims to contribute to these
efforts.

Here we build a theoretical model to generate steady state multipartite and
multilevel entanglement mediated solely by the environment. First we consider a
general open quantum system coupled to a bath and then we examine the
equilibrium reduced density matrix of non-interacting qudits through the
analytical polaron treatment, complemented by a numerically exact path-integral
approach. Next, we demonstrate the use of the structure-factor-based and global
entanglement witnesses to compute entanglement in bipartite and multilevel
systems. For a non-interacting pair of qubits, we demonstrate that the
entanglement is maximised at finite amount of system-bath interaction, and the
peak values are temperature dependent.

Our results can be tested experimentally, as scattering experiments have already
proved to be useful in quantifying multipartite entanglement~\cite{Marty2014}.
Structure factors are macroscopic quantities measurable in various periodic
systems, e.g., Bragg scattering on ultracold atoms in optical lattices
\cite{Corcovilos2010} or from ion chains \cite{Macchiavello2013} and neutron
scattering from solid state systems. Furthermore, the structure factor,
particularly the dynamic structure factor, is a central observable quantity in
determining single-particle and collective electronic excitations in many-body
systems \cite{Pitarke2007, Pines1966}.
%% -----------------------------------------------------------------------------
%% FIGURE: SCHEMATICS
%% -----------------------------------------------------------------------------
\begin{figure}[b!]
    \includegraphics[width=84mm]{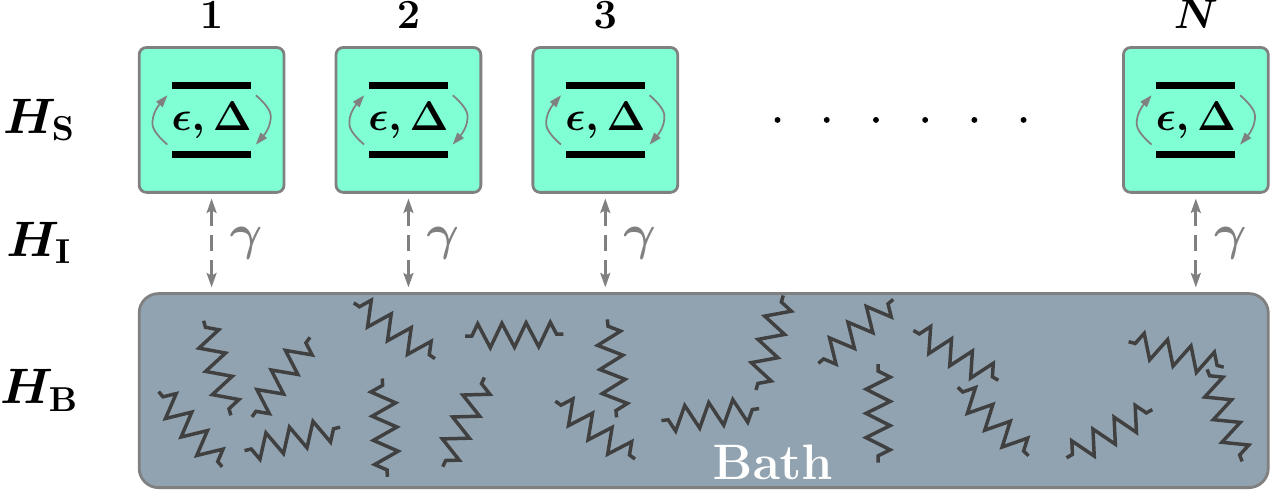}
    \caption{\label{fig:SchematicsOfEntireSystem}
             (Colour online) The schematics of the physical system consisting of
             $N$ $d$-level systems (qudits) and a bath of harmonic oscillators.
             The qudits do not directly interact with each other, but they are
             all coupled to the bath.
             }
\end{figure}
%% -----------------------------------------------------------------------------

%% =============================================================================
%% THEORY
%% =============================================================================
{\emph{Model} ---} We consider the collection of $N$ identical, non-interacting
qudits. The energy splittings between levels are assumed to be uniform and
denoted by $\epsilon$ while the tunnelling matrix element is $\Delta$. The bath
is modelled as an ensemble of harmonic oscillators with frequencies
$\omega_{n}$. The strength of coupling between a qudit and an oscillator with
angular frequency $\omega_{n}$ is denoted by $g_{n}$. Thus the total
system-plus-bath Hamiltonian is formally written as
$H_{\mathrm{tot}}=H_{\mathrm{S}}+H_{\mathrm{B}} + H_{{ \mathrm{I}}}$, where
$H_{\mathrm{S}}$, and $H_{\mathrm{B}}$ govern the internal dynamics of the
system and the bath, respectively, while $H_{\mathrm{I}}$ describes the coupling
between them. In detail
\begin{eqnarray*}
    H_{\mathrm{S}}
    &=&
    \frac{1}{2}
    \sum_{j=1}^{N}{\Bigl \lbrack
                    \epsilon s^{(j)}_{z} + \Delta s^{(j)}_{x}
                 \Bigr \rbrack
                },
    \\
    H_{\mathrm{B}}
    &=&
    \sum_{n}{\omega_{n}^{\phantom{\dagger}}
             b^{\dagger}_{n} b_{n}^{\phantom{\dagger}}
            },
    \\
    H_{\mathrm{I}}
    &=&
    \sum_{n}{\sum_{j=1}^{N}%
                 {g_{n}^{\phantom{\dagger}} s^{(j)}_{z}\!
                  \left ( b_{n}^{\dagger} + b_{n}^{\phantom{\dagger}} \right )
                 }},
\end{eqnarray*}
where $s_{x}$ and $s_{z}$ are the $x$ and $z$ components of a spin operator of a
system with spin $s$, and the $b_n$ are the bosonic bath operators. The number
of degrees of freedom of the bath is assumed to be much larger than that of the
qudits, thus according to the Poincar{\'e} recurrence theorem the time-scale on
which the bath feeds back the energy to the subsystems is enormous.

This separation of time-scales allows us to smooth the oscillator modes and
instead of $\lbrace \omega_{n} \rbrace$ we characterise the bath with a
continuous spectral density $J(\omega)$. For analytical calculations below, we
assume a super-Ohmic spectrum, $J(\omega) = \gamma (\frac{\omega}{\omega_{
\mathrm{c}}}) ^{3} e^{-\omega/\omega_{\mathrm{c}}}$, of exponential cut-off with
a fixed $\omega_{\mathrm{c}}$, and constant system-bath coupling strength,
$\gamma$, with the dimension of frequency. The reciprocal of the cut-off
frequency, $\tau \propto \tfrac{1}{\omega_{\mathrm{c}}}$ governs the relaxation
time of the bath. This form of the spectral density is commonly used in the
study of tunnelling effects in solid state systems \cite{Leggett1987}. Though
the analysis below utilises a super-Ohmic spectrum, numerical results with
other spectral densities, e.g., Ohmic and Lorentzian (see supplementary
information), show that the conclusions are applicable to a wide range of
systems. In order to capture the effects of strong qudit-bath coupling, we
employ the polaron transformation defined as $\po{H}_{\mathrm{tot}} = e^{P}
H_{\mathrm{tot}} e^{-P}$, where
\begin{equation*}
   P
    =
    2 \sum_{j}{s^{(j)}_{z}
             \sum_{n}{\frac{g_{n}}{\omega_n}
                      \left (
                          b_{n}^{\dagger} - b_{n}^{\phantom{^\dagger}}
                      \right )
                     }
            }.
\end{equation*}
The total Hamiltonian preserves its structure under this transformation:
$\po{H}_{\mathrm{tot}} = \po{H}_{\mathrm{S}} + \po{H}_{\mathrm{B}} +
\po{H}_{\mathrm{I}}$, where $\po{H}_{\mathrm{B}} = H_{\mathrm{B}}$ and
\begin{eqnarray*}
        \po{H}_{\mathrm{S}}
        &=&
        \frac{1}{2}
        \sum_{j=1}^{N}{\left \lbrack
                    \epsilon s^{(j)}_{z} +
                    \Delta_{\mathrm{R}}s^{(j)}_{x}
                 \right \rbrack
                }
        -
        \sum_{n}{\frac{g_{n}^{2}}{\omega_{n}}
                 \sum_{j,k}^{N}{s_{z}^{(j)} s_{z}^{(k)}}
                },
        \\
        \po{H}_{\mathrm{I}} \label{eq:4}
        &=&
        \sum_{j=1}^{N}%
            {\left \lbrack
                s_{x}^{(j)} V_{x} + s_{y}^{(j)} V_{y}
             \right \rbrack
            }.
\end{eqnarray*}
Several remarks are in order. First, the Hamiltonians are partitioned such that
the expectation value $\Tr_{\mathrm{SB}}{( \po{\rho} \po{H}_{\mathrm{I}})} =
\langle \po{H}_{\mathrm{I}} \rangle_{\mathrm{SB}} = 0$. Second, an effective
qudit-qudit interaction term, $\sim s_{z}^{(j)} s_{z}^{(k)}$, appears in
$\po{H}_S$, induced by the system-bath coupling. This term is the
source of the finite equilibrium entanglement in the system since the qudits
were originally assumed to be non-interacting. Third, the system-bath coupling
in the polaron picture, $\po{H}_{\mathrm{I}}$, assumes a form very different
from $H_{\mathrm{I}}$. All spin operators appear in a symmetric way and they are
coupled to {\emph{effective bath}} operators $V_{x}$, $V_{y}$. The explicit
expressions of these bath operators are immaterial for our current analysis but
for sake of transparency they are provided in the supplementary material.
Finally, the polaron transformation renormalises the tunnelling rate, $\Delta
\mapsto \Delta_{ \mathrm{R}}$, and it thus picks up a temperature and coupling
strength dependence.

The equilibrium reduced density matrix can formally be expressed
\cite{Weiss2012} as a partial trace
\begin{eqnarray*}
    \tilde{\rho}_{\mathrm{S}}
    =
    \frac{\Tr_{\mathrm{B}}{\!\left ( e^{- \beta \po{H}_{\mathrm{tot}}} \right )}}
         {\Tr_{\mathrm{SB}}{\!\left ( e^{-\beta \po{H}_{\mathrm{tot}}} \right )}}.
\end{eqnarray*}
Expanding $\tilde{\rho}_S$ and taking into account that $\langle \po{H}_{
\mathrm{I}} \rangle_{\mathrm{SB}}=0$ guarantees that in perturbative expansion
the leading order term for $\po{\rho}_{\mathrm{S}}$ is of second order in the
interaction Hamiltonian. The equilibrium state of the system can be
approximated~\cite{Lee2012} as $\po{\rho}_{\mathrm{S}} \approx \po{\rho}_{
\mathrm{S}, 0} + \po{\rho}_{\mathrm{S}, 2}$ where
\begin{align*}
    \po{\rho}_{\mathrm{S},0}
    &=
    Z_{\mathrm{S}, 0}^{-1} \, e^{-\beta \po{H}_{\mathrm{S}}}
    \\
    \po{\rho}_{\mathrm{S}, 2}
    &=
    Z_{\mathrm{S}, 0}^{-2}
    \left \lbrack
        A Z_{\mathrm{S}, 0}^{\phantom{-1}} -
        Z_{\mathrm{S}, 2}^{\phantom{-1}}\, e^{-\beta \tilde{H}_{\mathrm{S}}}
    \right \rbrack.
\end{align*}
The leading $\po{\rho}_{\mathrm{S},0}$ term is the expected canonical
equilibrium distribution in the polaron picture with $Z_{\mathrm{S}, 0}
= \Tr_{\mathrm{S}}{\!(e^{-\beta \po{H}_{\mathrm{S}}})}$, while the correction
term, $\po{\rho}_{\mathrm{S},2}$, picks up some of the correlation between the
multi-level subsystems and the bath via
\begin{equation*}
    A
    =
    \sum_{\ell}%
        {\int_{0}^{\beta}%
             {\!d\tau
              \int_{0}^{\tau}%
                  {\!d\tau' C_{\ell\ell}(\tau - \tau')
                   e^{-\beta \po{H}_{\mathrm{S}}}
                   S_{\ell}(\tau) S_{\ell}(\tau')}
              }
        },
\end{equation*}
where $S_{\ell} = \sum_{j} s^{(j)}_{\ell}$ and $Z_{\mathrm{S}, 2} = \Tr_{
\mathrm{S}}{\!(A)}$. Even the zeroth order term captures some correlation
through the transformed system Hamiltonian. The exact expressions of
$C_{\ell\ell}(\tau) = \langle V_{\ell}(\tau) V_{\ell} (\tau')
\rangle_{\po{H}_\mathrm{B}}$ are in the supplementary information.

It is useful at this point to analyse the behaviour of perturbation theory at
strong coupling in the polaron frame. As $\gamma \mapsto \infty$ the system
becomes incoherent since the coherent tunnelling element, $\Delta_\mathrm{R}$
vanishes. Simultaneously the correlation functions, $C_{\ell
\ell}(\mathrm{\tau})$, and hence the second-order correction to the reduced
density matrix $\po{\rho}_{\mathrm{S},2}$ vanishes. Therefore the equilibrium
density matrix at strong $\gamma$ is dominated by the bath induced coupling
term, $\sigma_{z}^{(j)} \sigma_{z}^{(k)}$,  e.g., for qubits
$\po{\rho}_{\mathrm{S}} \propto \exp(- \beta \po{H}_{\mathrm{S}})$, where
$\po{H}_{\mathrm{S}}$ is a diagonal matrix, since it is proportional to
$\sigma_z$. Therefore $\po{\rho}_{\mathrm{S}}$ is also diagonal. This result,
namely that $\po{\rho}_{\mathrm{S}}$ follows the canonical Gibbs' distribution,
at least in the strong coupling regime, is in agreement with \cite{Johnson2014}.
In the strong coupling (and limit of weak coupling) the density matrix can be
factored as a product of single particle density matrices  so this qubit system
can be considered as a classical simulator in the language of
\cite{Johnson2014}. It is only in the intermediate coupling regime, in the
presence of entanglement, that the system provides a quantum resource.

%% =============================================================================
%% Entanglement
%% =============================================================================
{\emph{Entanglement witness} ---} The Hahn--Banach separation theorem guarantees
that for any entangled state there is an entanglement witness~\footnote{In the
physics literature the concept of entanglement witness is predominantly defined
as a functional, $W$, of the density matrix, $\rho$, such that $\Tr{\!(W\!\rho)}
<0$ if $\rho$ is entangled, and $\Tr{\!(W\!\rho)}>0$ if $\rho$ is separable.}.
However, currently there is no universal measure of entanglement for an
arbitrary number of subsystems each possessing multiple levels, rather there are
a plethora of functionals which are useful in specific situations. Unlike qudit
and multipartite systems, bipartite qubit systems can be classified by a single
measure~\cite{Vedral1997}. This fact shows that bipartite qubit systems are not
typical, rather the exceptions.

In order to quantify bipartite qubit entanglement, we adopt the standard measure
of entanglement of formation (EoF) \cite{Wootters1998}. However, detecting
multipartite entanglement for a mixed state is more complicated
\cite{Krammer2009}, and we rely on the structure factor as a precursor of an
entanglement witness
\begin{equation*}
    {\mathfrak{S}}_{\ell \ell'}(k)
    =
    \sum_{i<j} e^{\iu k (r_j -r_i)}
    \left \langle s^{(i)}_{\ell} s^{(j)}_{\ell'} \right \rangle,
\end{equation*}
where $\ell$, $\ell'$ run over the set $\lbrace x, y, z \rbrace$, $k$ is the
wavenumber, and $r_{i}$, $r_{j}$ are the positions of spins $i$ and $j$.
The distance between two adjacent qubits is normalised to unity \footnote{The
physical distance of two qubits is arbitrary. If the qubits are at equidistant
locations with distance $a$ then a structure factor ${\mathfrak{S}}_{\ell \ell'}
(k, 1)$ calculated with unit distance and another structure factor
${\mathfrak{S}}_{\ell \ell'}(k, a)$ calculated with a different distance, $a$,
are related to each other as ${\mathfrak{S}}_{\ell \ell'}(k, 1) =
{\mathfrak{S}}_{\ell \ell'} (\frac{k}{a}, a)$}. The entanglement witness, based
on ${\mathfrak{S}}_{\ell \ell'}(k)$, is defined as $W(k) = \mathbb{1} -
\Sigma(k)$, where $\Sigma(k) = \bar{\Sigma}(k) + \bar{\Sigma}(-k)$, and
\begin{eqnarray*}
    \bar{\Sigma}(k)
    =
    \frac{1}{2 {N \choose 2}}
    \Bigl \lbrack
        c_{x} {\mathfrak{S}}_{xx}(k) +
        c_{y} {\mathfrak{S}}_{yy}(k) +
        {c_{z}}{\mathfrak{S}}_{zz}(k)
    \Bigr \rbrack.
\end{eqnarray*}
It has been proven~\cite{Krammer2009} that $W$ detects multipartite entanglement
of a mixed state whenever $\left \langle W \right \rangle = \Tr{\!\left ( W
\!\rho \right )} < 0$, or equivalently $\langle \Sigma \rangle > 1$. In general
the coefficients $c_{\ell}$ can be chosen arbitrarily by the observer provided
all $c_{\ell}$ are real and $\abs{c_{\ell}} \leq 1$. Below we use $(c_{x},
c_{y}, c_{z}) = (1, -1,1)$. In the case of qudit systems we use a convenient
lower bound~\cite{Tiranov2017} of the entanglement of formation
\begin{eqnarray*}
    \mathrm{EoF}
    \geq
    \mathrm{EoF}_{\mathrm{lb}}
    = -\log_{2}{\!\left ( 1- \tfrac{1}{2} R^{2} \right )},
\end{eqnarray*}
where
\begin{equation*}
    R
    =
    \frac{2}{\sqrt{\abs{C}}}
    \sum_{\substack{(j, k) \in C \\ j<k}}%
        {\!\!\Bigl \lbrack
         \abs{\left \langle
                  j, j \middle \vert \rho \middle\vert k, k
              \right \rangle} -
        \!\sqrt{\langle j , k|\rho|j,k\rangle \langle k,j|\rho|k,j\rangle}
        \Bigr \rbrack},
\end{equation*}
with $C$ being the set containing all pairs of indices $(j, k)$, while $\abs{C}$
denotes its cardinality.

%% =============================================================================
%% RESULTS
%% =============================================================================
{\emph{Results} ---} We first revisit the two-qubit case, i.e., $s_{\ell}
\mapsto \tfrac{1}{2} \sigma_{\ell}$, and while this system has been analysed
before \cite{Braun2002, McCutcheon2009} we expand those analyses here. We use an
imaginary time path-integral technique for calculating the reduced density
matrix for the bare system, while for the dressed particles the previously
described polaron transformation and second order approximation is employed
since it yields accurate equilibrium density matrices for a wide range of
parameters~\cite{Lee2012}, except for simultaneously low $\omega_{c}$ and
temperatures.

In the weak coupling regime, increasing the bath mediated interaction between
qubits enhances the induced entanglement. On the other hand, entanglement
decreases exponentially as a function of $\gamma$ in the strong coupling regime.
The interaction term, $\sim \sigma_{z}^{(j)} \sigma_{z}^{(k)}$ in $\tilde{H}_{
\mathrm{S}}$, dominates in the strong coupling limits, and the resulting $\po{
\rho}_{\mathrm{S}}$ is localized (i.e., diagonal) in the $\sigma_{z}$ basis with
zero entanglement. The scaling limits the entanglement in both weak and strong
coupling regimes and gives rise to the maximal entanglement at intermediate
$\gamma$. The non-monotonic behavior of EoF as a function of $\gamma$ may be
reminiscent of the ``stochastic resonance'' phenomenon reported by
Huelga~\cite{Huelga2007} and Lee~\cite{Lee2011}. Unlike the coupled quantum
many-body systems considered in these earlier works, the qubits here are not
coupled directly, the entanglement is solely induced by the interaction with the
common reservoir.

It should be noted that $\po{\rho}_{\mathrm{S}}$ is quite different from
$\rho_{\mathrm{S}}$. The transform describes a system ``dressed'' by the polaron
cloud due to the system-bath interaction. However, this effective system is
similar to those typically found in solid state quantum dots and BCS pairs. It
is interesting to compare the amount of entanglement between two dressed qubits
with that between two bare qubits. While the analytical polaron method is
capable of exploring the entire range of dissipation strength and temperature,
we need to employ the numerically exact imaginary time path-integral technique
\cite{Moix2012} to obtain $\rho_S$. Given the high computational cost of the
path-integral technique, we restricted our analysis to small systems, at
moderate $\gamma$ and temperature values.
%% -----------------------------------------------------------------------------
%% FIGURE: TWO QUBITS
%% -----------------------------------------------------------------------------
\begin{figure}[b!]  % FIGURE: Two qubit entanglement plot
    \includegraphics[width=84mm]{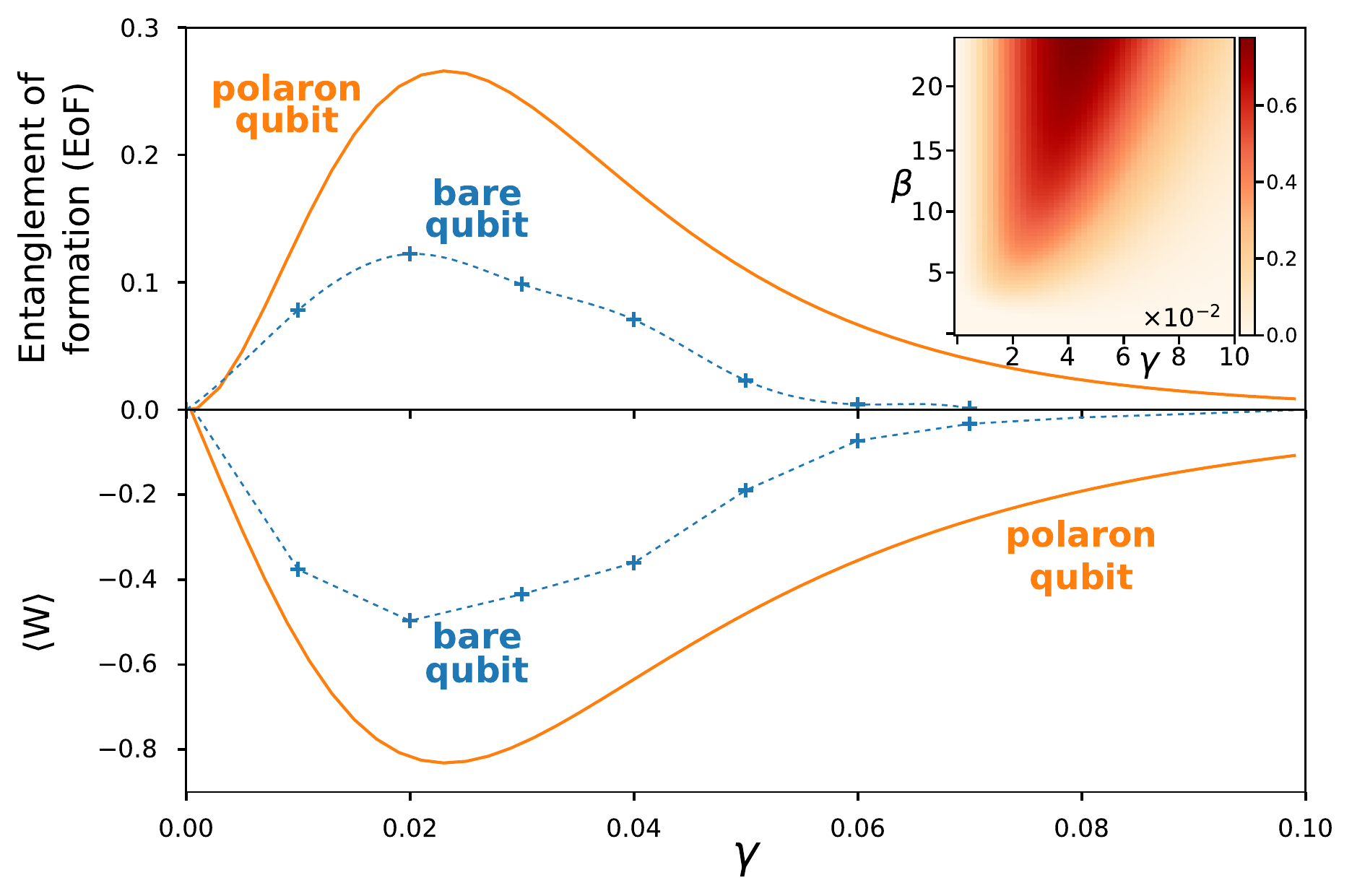}
    \caption{\label{fig:TwoQubitsCompareEntanglements_Heatmap}
             (Colour online) Two qubits: EoF and $\left\langle W \right
             \rangle$ are shown for two bare- and polaron qubits at $\beta=5$.
             The dashed lines are only to guide the eye. Inset provides the EoF
             between two polaron qubits as a heatmap for varying interaction
             strength, $\gamma$, and inverse temperature, $\beta$.
            }
\end{figure}

The EoF between the qubits as a function of dissipation strength, $\gamma$, is
shown in Fig.~\ref{fig:TwoQubitsCompareEntanglements_Heatmap}. The dependence of
entanglement on $\gamma$ between two bare qubits is similar to that of polaron
qubits, albeit at a lower value. The difference of EoF between dressed qubits
and bare qubits is a consequence of the system-phonon and phonon-phonon
entanglement, captured through the polaron transformation but not in the
calculation of entanglement between the bare qubits. The polaron transformation
dresses the qubit with phonons in the bath and correlations in this phonon gas
then contribute to the polaron entanglement. This result does not contradict our
assumption of Markovian (memory less) time-evolution, rather it reflects the
fact that within the environment equilibrium correlation has built up.

Fig.~\ref{fig:TwoQubitsCompareEntanglements_Heatmap} also illustrates the
structure-factor-based entanglement witness, $\langle W \rangle$, for the
two-qubit system. The negativity of $\langle W \rangle$ is proportional to the
entanglement of formation. Of course $\langle \Sigma \rangle$ could also be used
as an approximate measure of the entanglement. The advantage of both $\Sigma$
and $W$, is that they are both directly related to the structure factor,
${\mathfrak{S}}_{\ell \ell'}(k)$ and one can thus connect to the wealth of
knowledge in designing scattering experiments in order to interrogate the
coupled system. Furthermore, adjusting the constants $c_{\ell}$'s one may
explore different ``pockets'' of the high-dimensional Hilbert
space~\cite{Vedral2014} and can further classify the entangled state.

The inset of Fig.~\ref{fig:TwoQubitsCompareEntanglements_Heatmap} shows the
EoF between two polaron qubits over a large range of $\gamma$ and inverse
temperature, $\beta$. It is observed that the bath mediated equilibrium
entanglement is also sensitive to temperature, in addition to $\gamma$. At high
temperature, all density matrices approach a totally mixed state since all
eigenstates are equally populated, thus $\po{\rho}_{\mathrm{S}} \rightarrow
\mathbb{1}$ and $\po{\rho}_{\mathrm{S}, 0} \rightarrow \mathbb{1}$ as $A
\rightarrow 0$. As a result, entanglement is destroyed at high temperatures, as
expected. However, at lower temperatures the entanglement of dressed qubits
develops and may become significant ($\sim 0.8$).

\begin{figure}[b!]
    \includegraphics[width=84mm]%
                    {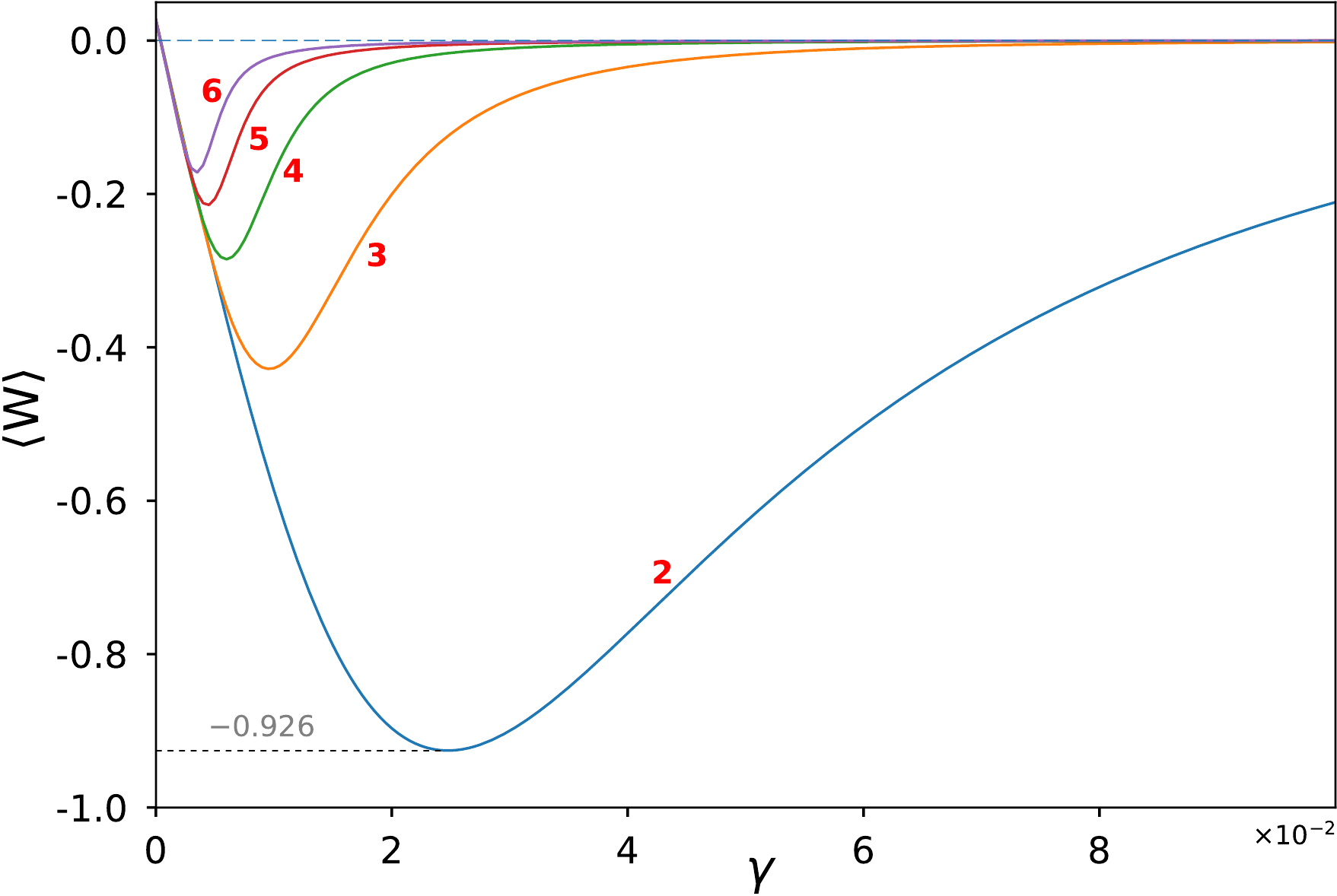}
     \includegraphics[width=84mm]%
                     {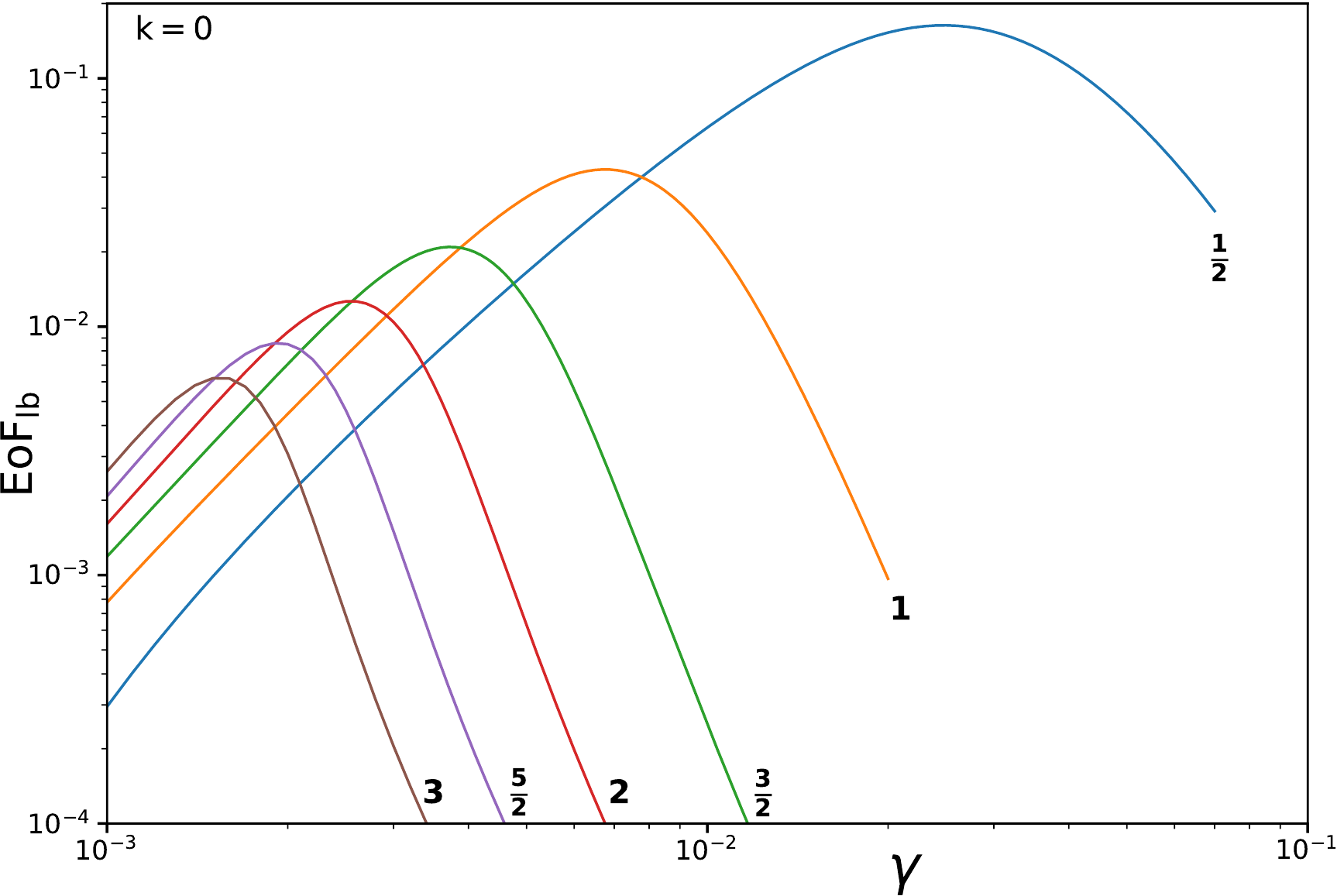}
    \caption{\label{fig:CompareStructureFactors}
             (Colour online) Entanglement witnesses, $\langle W \rangle$ for
             multipartite qubits for $N=2$, 3, 4, 5, and 6 (main figure). Other
             parameter values are $\beta=5$, $k=0$, $(c_{x}, c_{y}, c_{z}) = (1,
             -1, 1)$, $\epsilon=0$, $\Delta=1$, and $\omega_{c}=3$. Bottom
             figure shows $\mathrm{EoF}_{\mathrm{lb} }$ for bipartite qubits and
             qudits with total spin values from $s = \tfrac{1}{2}$, up to $3$ in
             steps of $\tfrac{1}{2}$. Notice the log-log scale.
            }
\end{figure}

Following the illustrative two-qubit case, we next explore the possibility
of observing bath induced multipartite entanglement. Let us consider a linear
chain of up to six noninteracting qubits ($N=6$) coupled to a common harmonic
bath. Fig.~\ref{fig:CompareStructureFactors} demonstrates the behaviour of
$\langle W \rangle$ for different system sizes, including $N=2$--6 qubits (red
bold labels). As seen, increasing the system size by one additional qubit causes
a significant decrease in optimal peak of $\langle W \rangle$ from $-0.926$ to
around $-0.4$ with the same noise level. Increasing the system size further, we
again observe a lower optimum peak for $\langle W \rangle$. Measuring $\langle W
\rangle$ for systems with $5$ and $6$ qubits shows that system size has a direct
effect on the correlation and the maximal entanglement moves to lower ranges in
the coupling for larger systems.
%% -----------------------------------------------------------------------------
%% FIGURE: ENTANGLEMENT WITNESSES
%% -----------------------------------------------------------------------------

%% -----------------------------------------------------------------------------
%% FIGURE: SIX QUBITS
%% -----------------------------------------------------------------------------
\begin{figure}[b!]
    \includegraphics[width=84mm]{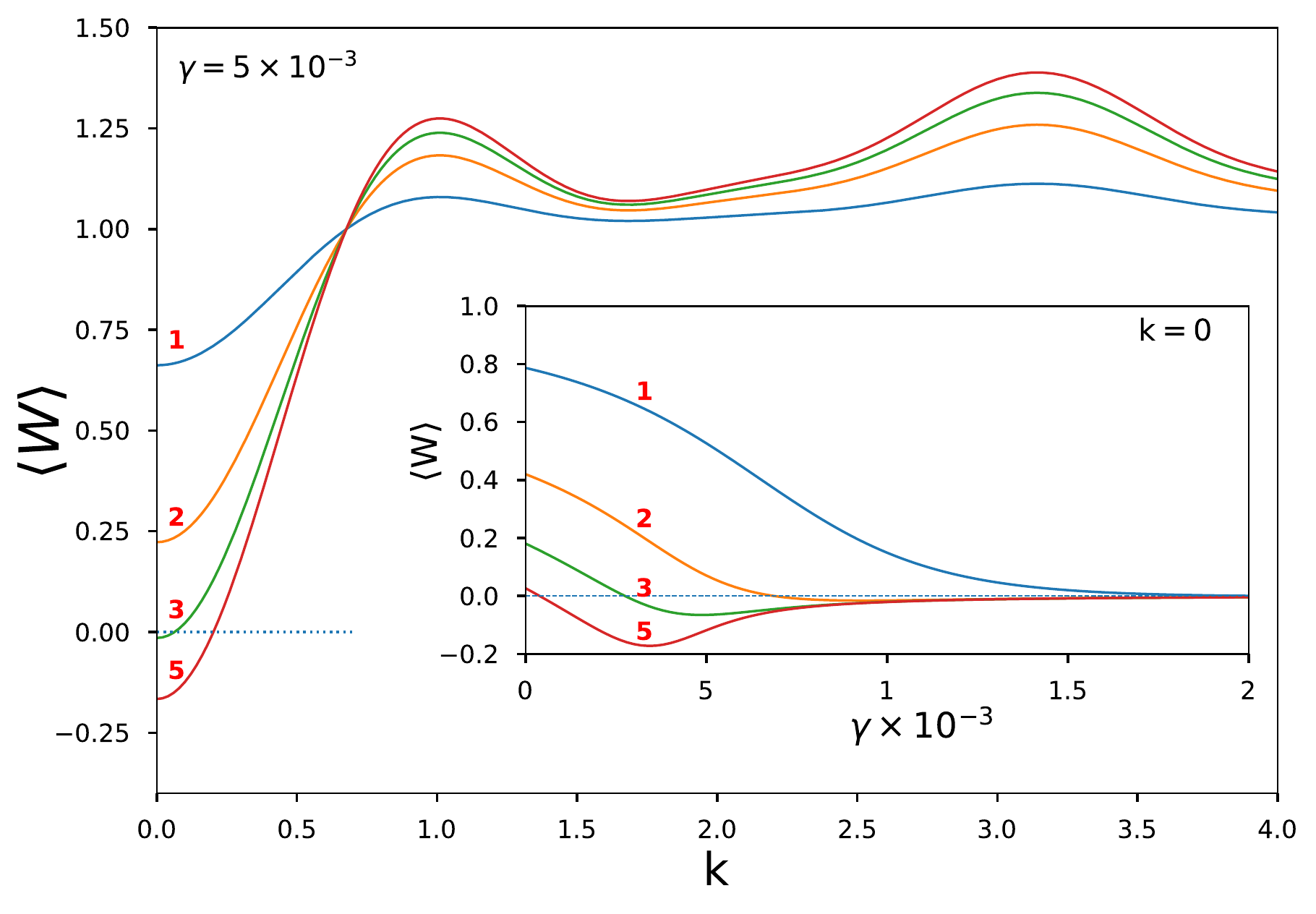}
    \includegraphics[width=84mm]{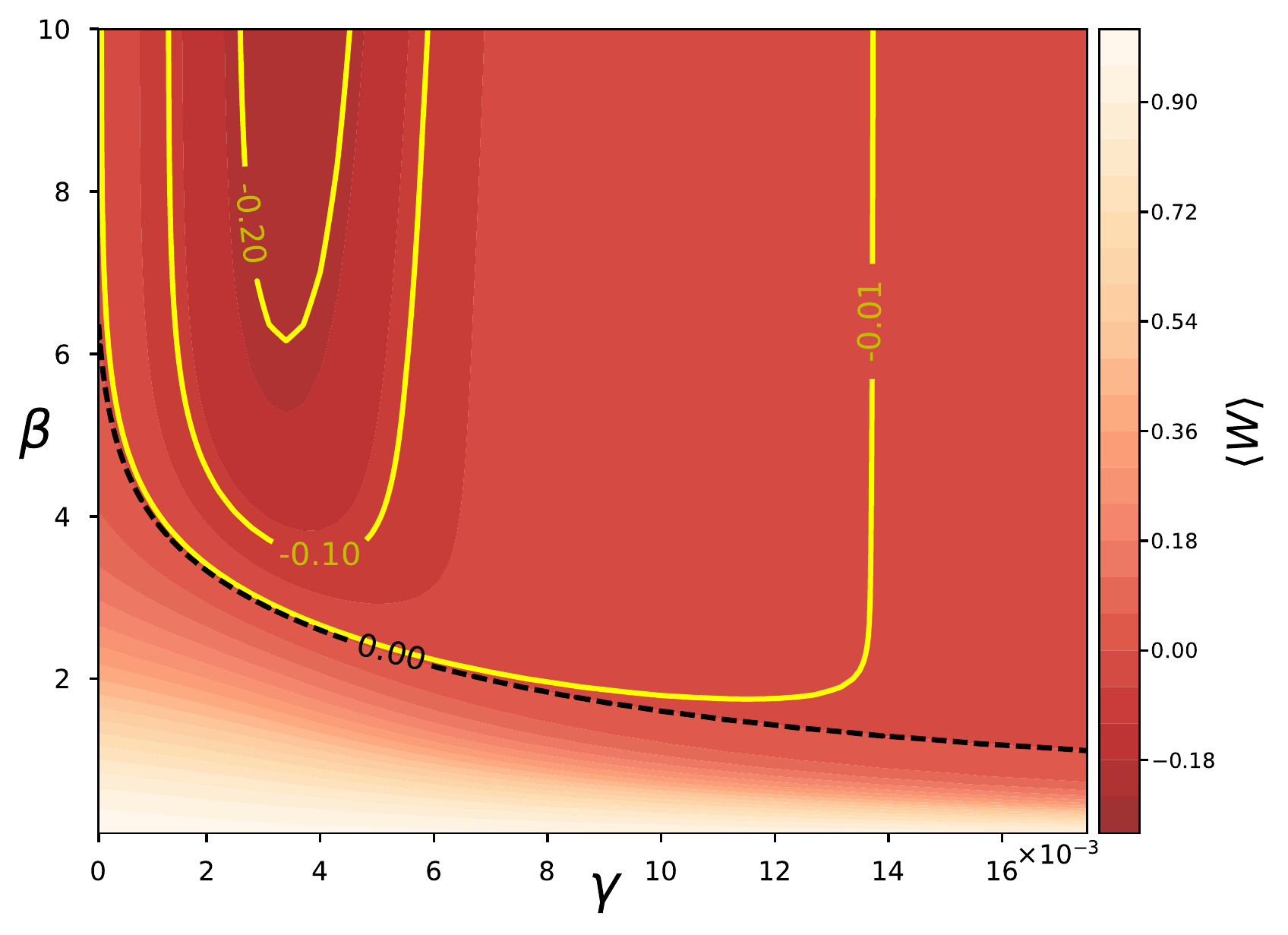}
    \caption{\label{fig:SixQubits_StructureFactor}
             (Colour online) Six qubits. (Top) Entanglement witness $\langle W
             \rangle$ is plotted as a function of $k$ with fixed interaction
             strength $\gamma = 5 \times 10^{-3}$. The inset depicts the
             dependence of $\langle W \rangle$ on $\gamma$ for the fixed $k=0$
             at different temperatures (red labels). (Bottom) Heatmap
             representation of $\left \langle W \right \rangle $ as a
             function of $\beta$ and $\gamma$ for $k=0$. The inner dark-red
             region denotes the parameter space where multipartite entanglement
             could be detected for $\left \langle W \right \rangle < 0$. The
             dashed black line separates the positive and negative regions of
             $\left \langle W \right \rangle$.
            }
\end{figure}

The inset of the top panel of Fig.~\ref{fig:SixQubits_StructureFactor} displays
the entanglement witness, $\langle W \rangle$, as a function of $\gamma$ at
different temperatures as indicated by the red, bold numbers which are the
dimensionless values of $\beta$. Similar to the bipartite case, it is shown that
multipartite entanglement, i.e., $\langle W \rangle < 0$, can be generated at
intermediate dissipation strength, $\gamma \approx 5 \times 10^{-3}$, provided
the temperature is low enough. As temperature increases, thermal fluctuations
destroy the equilibrium multipartite entanglement as the system reduced density
matrix approaches a totally mixed state. Turning to the main graph in the top
panel of Fig.~\ref{fig:SixQubits_StructureFactor}, a negative $\langle W
\rangle$ can be seen at low wavenumbers for low temperatures. This implies the
existence of multipartite entanglement occurring around $k \cong 0$ for $\beta >
3$.

Finally, the lower heatmap of Fig.~\ref{fig:SixQubits_StructureFactor} displays
$\langle W \rangle$ as a function of both $\beta$ and $\gamma$. Dashed black
line separates two distinct regimes of multipartite entanglement. It seems
worthwhile mentioning the generic \textsc{u}-shape of the solid lines,
indicating that multipartite entanglement develops only within a range of
$\gamma$ values and below a critical temperature value. While the latter
feature, the existence of a critical temperature is expected, the strong
dependence on $\gamma$ is --we believe-- novel. The steep slope of the solid
curve in terms of $\gamma$ suggests that weak coupling to the bath cannot
efficiently relay entanglement and the qubits evolve in time more or less
independently of each other. On the other hand, strong coupling with the bath
connects the qubit with numerous phonon modes, leading to decoherence and the
inhibition of any transmission of entanglement between the qubits.

From here onwards we focus on multilevel systems. One may extend the results
above to higher spin values by using appropriate spin operators as the formalism
has already been presented in this general form. For example, the form of the
Hamiltonian for a spin-1 system (qutrits) is similar to that of the
spin-$\tfrac{1}{2}$ system, with the main difference being the occurrence of the
anisotropic term, $\sim s_{z}^{2}$ in $\po{H}_{\mathrm{S}}$. This term
contributes only a constant in the case of $s_{z} = \frac{1}{2} \sigma_{z}$. The
polaron transformation can still be defined in this higher dimensional space. In
order to apply the transform to $H_{\mathrm{tot}}$, the Baker-Campbell-Hausdorff
formula~\cite{Hall2003} is needed, but this does not pose any serious limitation
on the calculation.

The bottom panel of Fig.~\ref{fig:CompareStructureFactors} compares
$\mathrm{EoF}_{\mathrm{lb}}$ for bipartite qudit systems from $s=\tfrac{1}{2}$
up to $s=3$ in steps of $\tfrac{1}{2}$ (red bold labels). Notice the log-log
scale of the axes. Increasing the dimensionality of a subsystem manifests itself
in more brittle entanglement such that at the same noise-level in the bath lower
entanglement can be achieved. Therefore the maximal entanglement shift towards
weaker interactions by increasing spin dimensions.

\begin{figure}[h!]
    \includegraphics[width=84mm]{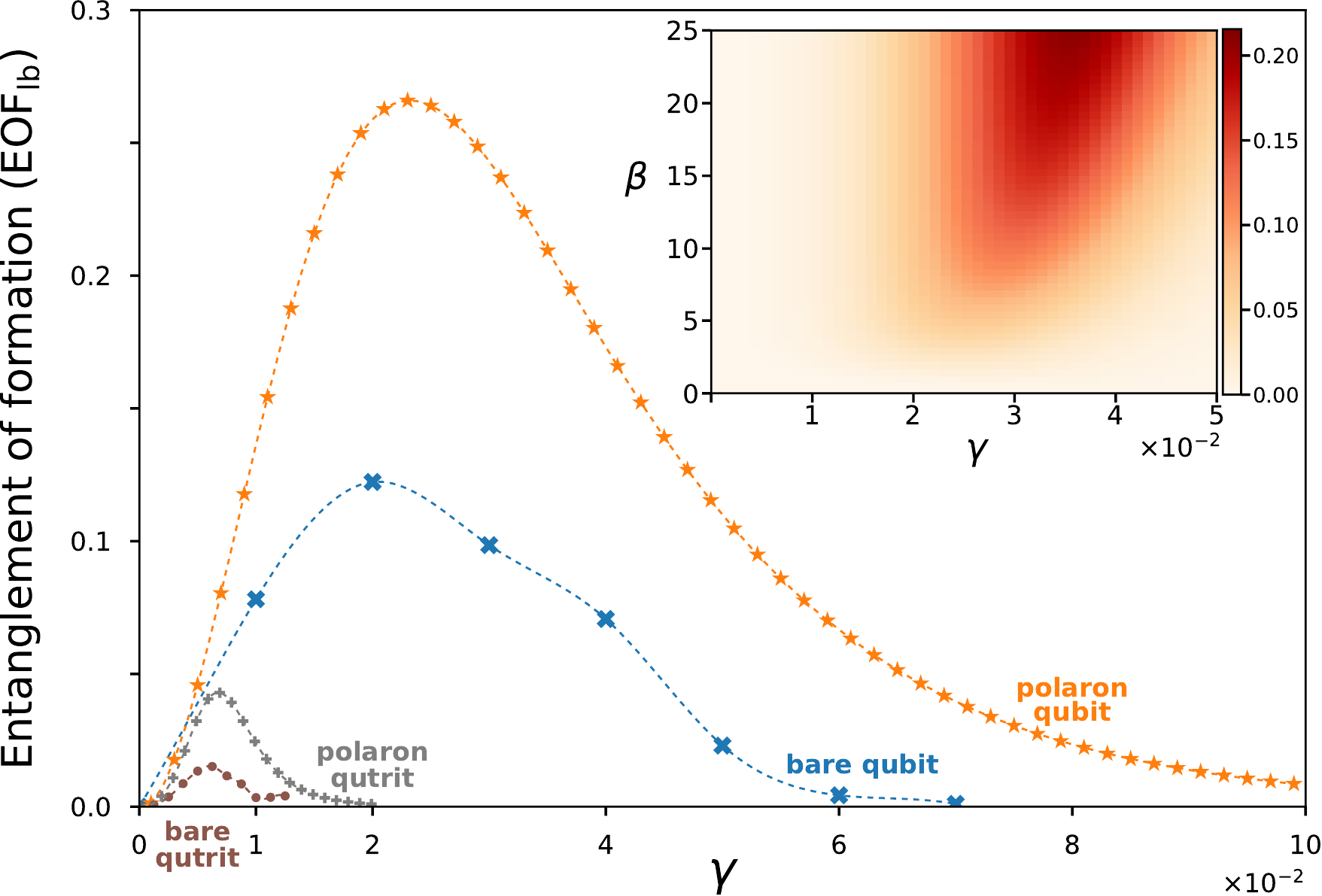}
    \caption{\label{fig:CompareQubitsAndQutrits}
             (Colour online) Figure focuses on the case of $s=1$ and
             compares the two entanglement measures for two qubits and two
             qutrits. Both their bare and polaron-dressed measures are given. It
             is apparent that entanglement is reduced significantly for higher
             spins. In the inset the lower bound of entanglement of formation
             ($\mathrm{EoF}_\mathrm{lb}$) is plotted for two polaron qutrits as
             a heatmap over a range of system-bath coupling, $\gamma$, and
             inverse temperature, $\beta$.
            }
\end{figure}

Now looking at Fig.~\ref{fig:CompareQubitsAndQutrits}, we see that the
entanglement of formation exhibits a stochastic resonance behaviour in qutrit
systems similar to that in the qubit system, but to a more limited extent.
%% -----------------------------------------------------------------------------
%% FIGURE: TWO QUTRITS
%% -----------------------------------------------------------------------------

Moreover, by comparing the curves of qubits and qutrits in the main panel of
Fig.~\ref{fig:CompareQubitsAndQutrits}, we may conclude that observable
entanglement in qutrits system can occur only at lower temperatures than in
qubit systems. This is expected as the temperature is a proxy for the noise in
the phonon-bath. Thus increasing temperature rises noise-levels which destroy
correlation. The stochastic resonance thus happens at lower noise levels for the
qutrit systems. In summary, the maximal entanglement in both cases depends on
temperature and happens at moderate system-bath coupling strengths.

Concluding, we have demonstrated that steady state multipartite and multilevel
entanglement can be induced solely through interactions with a common reservoir.
The multipartite and multilevel entanglement are experimentally accessible and
we propose that an analysis of the system's structure factors could be used for
their detection. Structure factors, both static and dynamic, are commonly used
in solid-state physics, and are suited to the capture of correlational structure
in few-body systems. They are measurable in scattering experiments, e.g., via
neutron scattering in condensed-matter systems, or via light scattering off
optical lattices.

We reiterate that the analysis presented here focuses on a bath with a
super-Ohmic spectrum for analytical treatment of the polaron perturbation
technique. Numerical results with other spectral densities (Ohmic and
Lorentzian) show that our conclusions are essentially independent of the precise
form of spectral density and applicable to a wide range of systems.

%% =============================================================================
%% CONCLUSIONS
%% =============================================================================

It is clear that the spin-boson models are an important tool in understanding
imperfections in quantum gate operation. The results presented above make some
first steps in characterising quantitatively steady-state multipartite and
multi-dimensional entanglement, thereby helping provide an additional resource
for quantum information applications. We envision that entanglement measures
based on the structure factor could become a good tool to probe and quantify
multipartite qudit entanglement and simultaneously test setups with controllable
dissipation, e.g., trapped ions and superconducting qubits.

Fermions (qudits) in optical lattices seem particularly suitable candidates for
investigation as these systems are highly controllable, e.g., the potential
depth and external magnetic field can control $\epsilon$ and $\Delta$, and how
strongly these atoms interact with each others. If a thermal cloud of atoms (the
bath) were trapped in a different potential and brought in contact with the
strongly trapped atoms in the lattice, their interaction, $\gamma$, could also
be tuned. Then one may potentially interrogate the trapped atoms via scattering
\cite{Marty2014, Landig2015}, and deduce the structure factor to quantify
entanglement.

{\emph{Acknowledgements} ---} We thank our funding agencies, particularly the NZ
Tertiary Education Commission, the National Research Foundation Singapore and
the Ministry of Education Singapore. One of us (DAWH) would like to thank Merton
College, Oxford, where he is a Visiting Research Fellow, for their generous
hospitality.

%%%%%%%%%%%%%%%%%%%%%%%%%%%%%%%%%%%%%%%%%%%%%%%%%%%%%%%%%%%%%%%%%%%%%%%%%%%%%%%
%% =============================================================================
%% APPENDIX A
%% =============================================================================
\noindent
{\emph{\bf Appendix A: Details of multipartite qubit system}}
We now discuss in a little more detail the qubit system. The total Hamiltonian
for $N$ qubits, $H_{\mathrm{tot}}
= H_{\mathrm{S}} + H_{\mathrm{B}} +  H_{\mathrm{SB}}$, in the open quantum
system formalism is given by
\begin{eqnarray*} % Total Hamiltonian
    H_{\mathrm{S}}
    &=&
    \frac{1}{2}
    \sum_{j}^{N}
        {\Bigl \lbrack
            \epsilon \sigma^{(j)}_{z} + \Delta \sigma^{(j)}_{x}
         \Bigr \rbrack
        },
    \\
    H_{\mathrm{B}}
    &=&
    \sum_{n}
        {\omega_{n}^{\phantom{\dagger}}
        b_{n}^{\dagger} b_{n}^{\phantom{\dagger}}}
    \\
    H_{\mathrm{SB}}
    &=&
    \sum_{nj}%
        {g_{n} \sigma^{(j)}_{z}
         \bigl ( b_{n}^{\dagger} + b_{n}^{\phantom{\dagger}} \bigr )},
\end{eqnarray*}
where $H_{\mathrm{S}}$, $H_{\mathrm{B}}$ and $H_{\mathrm{SB}}$ are the system,
bath and the system-bath coupling Hamiltonians, respectively. The system is
described by a group of $N$ non-interacting identical two-level systems with
energy splitting $\epsilon$ and tunnelling matrix element $\Delta$. The
corresponding polaron transformation is
\begin{equation*}
    P
    =
    \sum_{j}%
        {\sigma^{(j)}_{z}
         \sum_{n}
             {\frac{g_n}{\omega_n}(b_{n}^{\dagger} - b_{n}^{\phantom{\dagger}})}
        }.
\end{equation*}
Up to a constant we obtain $\po{H}_{\mathrm{tot}} = \po{H}_{\mathrm{S}} +
\po{H}_{\mathrm{B}} + \po{H}_{\mathrm{SB}}$ with
\begin{eqnarray*}
    \po{H}_{\mathrm{S}}
    &=&
    \sum_{j}%
        {\left \lbrack
            \frac{\epsilon}{2} \sigma_{z}^{(j)} +
            \frac{\Delta_{R}}{2} \sigma_{x}^{(j)}
         \right \rbrack }
    -
    \sum_n \frac{g_n^2}{\omega_n}  \sum_{jk}\sigma_z^{(j)} \sigma_z^{(k)},
    \\
    \po{H}_{\mathrm{B}}
    &=&
    H_{\mathrm{B}}
    \\
    \po{H}_{\mathrm{SB}}
    &=&
    \sum_{j} \Big[\sigma_x^{(j)} V_x + \sigma_y^{(j)} V_y \Big].
\end{eqnarray*}
For $\po{H}_{\mathrm{S}}$, the tunnelling element in $\po{H}_{\mathrm{S}}$ is
renormalized due to the system bath coupling,

\begin{equation*}
    \Delta_{\mathrm{R}}
    =
    \Delta
    \exp{\!\left (
            -\frac{2}{\pi}
            \int^{\infty}_{0}
                {
                 \frac{J(\omega)}{\omega^2}
                 \coth{\!\left ( \frac{\beta \omega}{2} \right )}\,
                 d\omega
                }
         \!\right )}.
\end{equation*}
%% -----------------------------------------------------------------------------
%% FIGURE: TWO QUBITS IN BATHS WITH DIFFERENT SPECTRAL DENSITIES
%% -----------------------------------------------------------------------------
\begin{figure}[h!]
    \includegraphics[width=84mm]{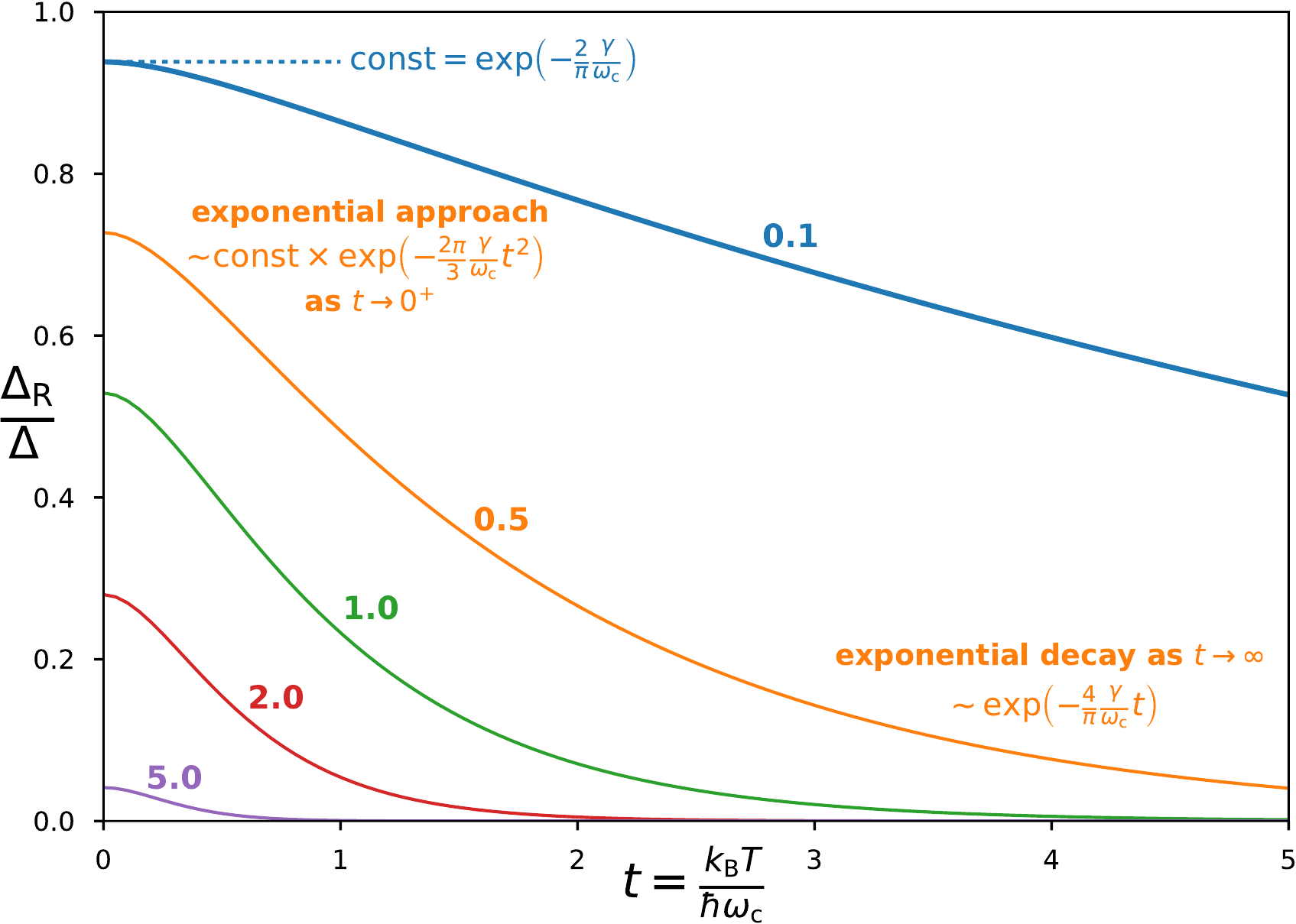}
    \caption{\label{fig:RenormalisingRation}
             (Colour online) The dimensionless ratio of $\Delta_{\mathrm{R}}/
             \Delta$ is plotted as a function of dimensionless temperature, $t =
             k_{\mathrm{B}}T/(\hbar \omega_{\mathrm{c}})$. The bold values
             close to each curves give the actual value of the coupling
             strength, $\gamma$. Asymptotic behaviour both at low- and
             high-temperatures are indicated. The $t \rightarrow 0^{+}$ limit
             also demonstrates that increasing $\gamma$ reduced the ratio
             exponentially fast, i.e, the tunnelling term in the polaron
             picture, $\Delta_{\mathrm{R}}$, quickly diminishes with
             stronger coupling.
            }
\end{figure}
In order to get some insight into the behaviour of the renormalised tunnelling
rate, $\Delta_{\mathrm{R}}$, one may evaluate this integral for a generic
super-Ohmic spectral density
\begin{equation*}
    J(\omega)
    =
    \gamma \,
    \left ( \!\frac{\omega}{\omega_{\mathrm{c}}} \right )^{\!k}
    \!\! e^{-\omega/\omega_{\mathrm{c}}}
\end{equation*}
where $k$ has to be bigger that 2 to mollify the singularity at $\omega=0$.
Substituting this expression into the integral one may arrive at
\begin{equation*}
    \ln{\!\left ( \frac{\Delta_{\mathrm{R}}}{\Delta} \right )}
    =
    - \frac{2}{\pi} \left ( \frac{\gamma}{\omega_{c}} \right ) \Gamma(k-1)
    \Bigl \lbrack 2 t^{k-1} \zeta(k-1, t) - 1 \Bigr \rbrack
\end{equation*}
where $\zeta(s,a)$ is the generalised Riemann-zeta function and $t =
k_{\mathrm{B}}T/(\hbar \omega_{\mathrm{c}})$ is a dimensionless temperature. Let
us choose $k=3$ in order to be consistent with the spectral density used in the
main body of the work and examine the low- and high-temperature asymptotic of
this expression. As $t \rightarrow 0^{+}$ the generalised Riemann-zeta function
develops a singularity, however, its pre-factor $t^{k-1}$ cancels that precisely
and we obtain
\begin{equation*}
    \frac{\Delta_{\mathrm{R}}}{\Delta}
    \cong
    \exp{\!\left (
            - \frac{2}{\pi}
            \left ( \frac{\gamma}{\omega_{\mathrm{c}}} \right )
            \Bigl \lbrack 1 + \frac{\pi^{2}}{3} t^{2} \Bigr \rbrack
         \right )}
    \rightarrow
    \exp{\!\left (
            - \frac{2}{\pi}
            \frac{\gamma}{\omega_{\mathrm{c}}}
         \right )}.
\end{equation*}
Thus the renormalised tunnelling term approaches a constant which depends on
the coupling strength and cut-off frequency. This result demonstrates that with
increasing system-bath coupling the tunnelling term vanishes exponentially. In
the opposite, high-temperature limit, the zeta function vanishes as
$(k-2)t^{2-k} + 2 t^{1-k}$, thus in our case ($\gamma>0$)
\begin{equation*}
    \frac{\Delta_{\mathrm{R}}}{\Delta}
    \cong
    \exp{\!\left ( -\frac{4}{\pi} \frac{\gamma}{\omega_{\mathrm{c}}} t \right )}
    \rightarrow
    0.
\end{equation*}
As anticipated, for decoupled systems, i.e., $\gamma =0$, the
tunnelling term is not affected by the bath in any way, thus $\Delta_{\mathrm{R}}
= \Delta$.

After analysing the renormalised tunnelling term in detail, let us return to
the formalism used for the two qubit case. As for the general spin case, here
the system-bath coupling also induces a qubit-qubit interaction term,
$\sigma_{z}^{(j)} \sigma_{z}^{(k)}$, which appears in $\po{H}_{\mathrm{S}}$.
The system bath coupling in the polaron picture, $\po{H}_{\mathrm{SB}}$, assumes
a form very different from $H_{\mathrm{SB}}$. The bath operators entering into
the system-bath coupling are
\begin{eqnarray*}
    V_{x} &=&  \frac{1}{4} \Delta \left  ( D_{+}^{} + D_{-}^{2} -2 B \right )
    \\
    V_{y} &=&  -\frac{i }{4} \Delta \left ( D_{-}^{2} - D_{+}^{2} \right ),
\end{eqnarray*}
with $D_{\pm} = \exp{\!\bigl ( \pm \sum_{n} \frac{g_{n}}{\omega_{n}} \left (
b_{n}^{\dagger} - b_{n}^{\phantom{\dagger}} \right ) \bigr )}$ and $\langle
D_{\pm}^2\rangle_{\mathrm{B}} = R$. The bath correlation functions,
$C_{\ell\ell'}(\tau) = \langle V_{\ell} (\tau) V_{\ell'} \rangle_{\mathrm{B}}$
are given by
\begin{eqnarray*}
    C_{xx}(\tau)
    &=&
    \frac{1}{4}
    \Delta_{R}^{2}
    \bigl \lbrack \cosh{(\phi(\tau))} - 1 \bigr \rbrack
    \\
    C_{yy}(\tau)
    &=&
    \frac{1}{4}\, \Delta_{R}^{2}
    \sinh{(\phi(\tau))},
\end{eqnarray*}
where
\begin{equation*}
    \phi(\tau)
    =
    \frac{4}{\pi}
    \int_{0}^{\infty}%
        {
         \frac{J(\omega)}{\omega^{2}}
         \frac{\cosh{\bigl ( \frac{1}{2} \, (\beta -2\tau)\, \omega \bigr )}}%
              {\sinh{\bigl ( \tfrac{1}{2} \beta \omega \bigr )}}\,
        d\omega
        }.
\end{equation*}
The diagonal correlation function, $C_{zz}(\tau)$, together with
cross-correlations, e.g., $C_{xz}(\tau)$, vanish identically
using the full polaron transformation described in the text.

%% =============================================================================
%% APPENDIX B
%% =============================================================================
\noindent
{\emph{\bf Appendix B: Ohmic and Lorentzian baths}}
Here we study the equilibrium entanglement induced by a common Ohmic or
Lorentzian bath. These baths are represented by the spectral densities
\begin{eqnarray*}
    J_{\text{Ohmic}}(\omega)
    &=&
    \gamma\, \omega\, \mathrm{e}^{-\omega/\omega_{c}}
    \\
    J_{\text{Lorentz}}(\omega)
    &=&
    \gamma \,\frac{\omega \,\omega_{c}}{\omega^{2} +\omega_c^{2}},
\end{eqnarray*}
respectively. Both of these densities behave as $J(\omega) \propto \omega$
for $\omega \ll 1$, thus an essential singularity appears at $\omega=0$ in the
expressions for $\Delta_{\mathrm{R}}$ and $\phi(\tau)$. Consequently, the
polaron method used in the main text is not applicable for these baths as it
suffers from nonphysical divergences~\cite{Weiss2012, Zwerger1983}. The integrals
in $\Delta_{\mathrm{R}}$ and $\phi(\tau)$ are all divergent for any non-zero
coupling strengths, thus the tunnelling element, $\Delta_{\mathrm{R}}$, is
always normalised to zero. Therefore, here we only present the numerical results
from the imaginary time path integral simulations.

In the spin-boson models, e.g., the one used in this work, where the bath is
harmonic, the trace over the bath degrees of freedom for $\po{\rho}_{\mathrm{S}}$
can be performed analytically and --within the path integral formalism-- we
arrive at the Feynman-Vernon influence functional~\cite{Feynman1963, Weiss2012}.
Employing the Hubbard-Stratonovich transformation and realising that the
influence functional can be unravelled by an auxiliary stochastic field, $\xi(
\tau)$, leads to a time-dependent Hamiltonian $H_{\mathrm{st}} \sim \tfrac{1}{2}
\bigl \lbrack \epsilon \sigma_{z} + \Delta \sigma_{x} \bigr \rbrack + \xi(\tau)
\sigma_{z}$ governing the imaginary time evolution. All of the effects of the
bath are accounted for by the coloured noise term, $\xi(\tau)$. The trace over
the bath now corresponds to averaging the imaginary time dynamics over
realisations of the noise. The primary benefit of this path integral approach is
that it provides the entire reduced density matrix from a single Monte Carlo
calculation and arbitrary spectral density, $J(\omega)$, may be used. In our
calculations, $10^{8} - 10^{11}$ samples achieved convergence. Further details
of the numerical implementation can be found in Refs.~\cite{Lee2012, Moix2012}.
%% -----------------------------------------------------------------------------
%% FIGURE: TWO QUBITS IN BATHS WITH DIFFERENT SPECTRAL DENSITIES
%% -----------------------------------------------------------------------------
\begin{figure}[t!]
    \includegraphics[width=84mm]%
                    {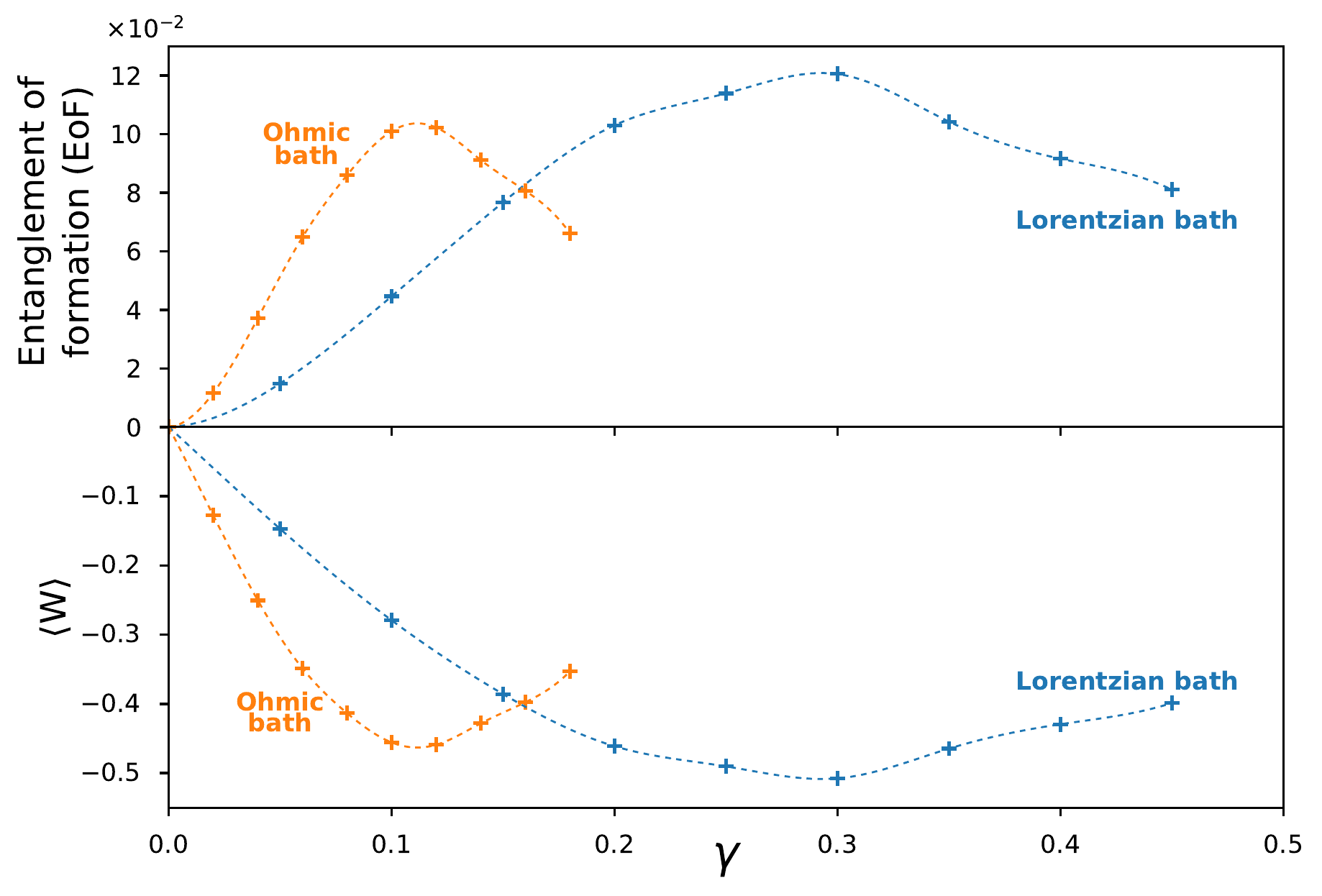}
    \caption{\label{fig:LorentzianAndOhmicMeasures}
             (Colour online) Entanglement measures, EoF and $\langle W \rangle$,
             are plotted as functions of the coupling strength, $\gamma$, for
             two qubits coupled to either an Ohmic (orange) or Lorentzian bath
             (blue). In both cases the inverse temperature is $\beta=10$, while
             the cut-off angular frequencies were $\omega_{\mathrm{c}} = 8$ for
             the Ohmic and $\omega_{\mathrm{c}} = 5$ for the Lorentzian bath.
             Note that the values of the ordinate in the top figure need to be
             multiplied by $\times 10^{-2}$.
            }
\end{figure}
Due to the high computational cost of this method, we only study the two qubit
case ($N=2$) for the weak and intermediate system-bath coupling regimes. In the
following, we use $\beta=10$, $\epsilon=0$ and $\Delta=1$. The entanglement of
formation and structure factor entanglement witness between two non-interacting
qubits in a common Ohmic bath and the same quantities for a Lorentzian bath are
shown in Fig.~\ref{fig:LorentzianAndOhmicMeasures}. It can be seen that the
features in both cases are qualitatively similar to those obtained using a
super-Ohmic spectral density in the main text. Therefore, the general
observations made in our main text should not be sensitive to the spectral
density of the bath.\\

%% =====================================================================================
%% APPENDIX C
%% =====================================================================================
\noindent
{\emph{\bf Appendix C: Scaling peak of entanglement and structure factor}}
%% FIGURE: PEAK ENTANGLEMENT
%% -----------------------------------------------------------------------------
Finally, we present an approximate power law scaling relation in the qudit
systems. Fig.~\ref{fig:Figure7_Supplementary_ScalingOfPeakEntanglement} shows
the scaling of peak entanglement for high dimensional bipartite and multipartite
systems. The $\mathrm{EoF}_{\mathrm{lb}}$ peaks obey a consistent power law,
although some fluctuation is seen in the peak of the structure-factor-based
entanglement measure for multipartite system. Relations vary between
approximately linear and a square root behaviour. An analytic treatment remains
the subject of further work.

\begin{figure}[h!]
    \includegraphics[width=84mm]%
                    {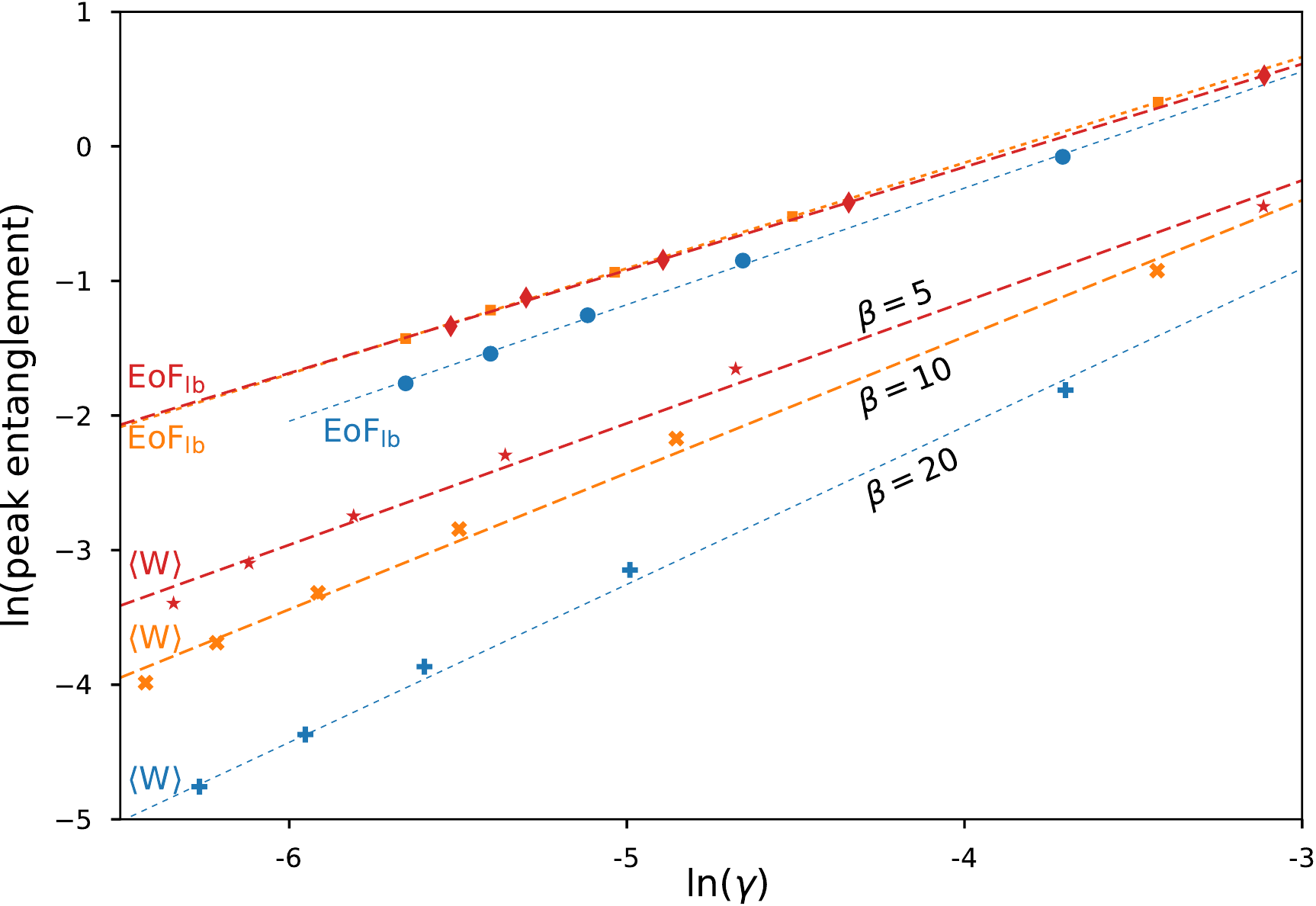}
    \caption{\label{fig:Figure6_Supplementary_ScalingOfPeakEntanglement}
             (Colour online) The value of peak entanglement measures,
             $\mathrm{EoF}_{\mathrm{lb}}$ and $\langle W \rangle$, as a function
             of interaction strength, $\gamma$, as a log-log plot together with
             linear fits. Powers vary between approximately 0.5 and 1.0. The
             inverse temperature values, $\beta=5$, 10 and 20, are indicated as
             appropriate.
            }
\end{figure}

%% =============================================================================
%% BIBLIOGRAPHY
%% =============================================================================
\bibliographystyle{apsrev4-1}
\bibliography{qubits_qutrits}

%% =============================================================================
%% AUTHOR CONTRIBUTIONS
%% =============================================================================
%\section{Author Contributions}
%    C.~K.~L., M.~S.~N., L.~C.~K and D.~A.~W.~H. undertook preliminary studies.
%    C.~K.~L., L.~C.~K conceived the simulation algorithms. C.~K.~L implemented
%    and together with M.~S.~N tested and ran all algorithms. The manuscript and
%    the illustrations were predominantly prepared by M.~S.~N., S.~D and
%    D.~A.~W.~H.

%% =============================================================================
%% ADDITIONAL INFORMATION
%% =============================================================================
% \section{Additional Information}
% {\textbf{Competing financial interests}}: The authors declare no competing
% interests.

\end{document}